\documentclass{moriond}

\usepackage{graphicx}
\usepackage{subfigure}
\usepackage[square,sort,comma,numbers,compress]{natbib}

\def\be{\begin{equation}}
\def\ee{\end{equation}}
\def\bea{\begin{eqnarray}}
\def\eea{\end{eqnarray}}

\newcommand{\Photo}{}

\begin{document}
\vspace*{4cm}
\title{PeV NEUTRINOS FROM ULTRA-HIGH-ENERGY COSMIC RAYS}

%\author{ A.B. AUTHOR }
\author{ G\"UNTER SIGL, ARJEN VAN VLIET }

\address{II. Institut f\"ur Theoretische Physik, Universit\"at Hamburg\\ Luruper Chaussee 149, 22761 Hamburg, Germany}

\maketitle\abstracts{
We investigate the possibility that the recently detected TeV-PeV
neutrino events by IceCube can originate from extragalactic
ultra-high-energy cosmic ray interactions with the cosmic microwave
background or the UV/optical/IR background. This is done by simulating
the propagation of the cosmic rays from their sources to the observer, including 
the production and propagation of secondary neutrinos and gamma rays.
For this purpose we use the publicly available 
simulation package CRPropa 2.0. We find that in all the scenarios
considered here the simulated neutrino flux level remains at least
one order of magnitude below the flux level indicated by the IceCube
events, thus showing it is difficult to interpret the IceCube events
in terms of a cosmogenic neutrino flux. 
}

\section{Introduction}

When ultra-high-energy cosmic rays (UHECRs) traverse the universe they
interact with extragalactic background light like the cosmic microwave 
background (CMB) and the UV/optical/IR background (IRB). One possible 
interaction with the CMB or IRB is photopion production. In this case 
the process $p+\gamma \rightarrow n+\pi^+$ will produce neutrinos from 
the decay of the neutron as well as from pion decay. When the UHECRs 
are nuclei instead of protons these nuclei can be photodisintegrated by 
photons from the CMB or IRB. In this way single neutrons can be 
separated from the nuclei, which will again decay and produce 
neutrinos. The nuclei themselves could also become unstable in this way 
and emit neutrinos in their decay. Here we investigate whether the 
recently observed neutrinos by IceCube with energies between 30~TeV and
2~PeV~\cite{Aartsen:2014gkd} could have originated from such 
interactions. This proceeding is based on Ref.~\cite{Roulet:2012rv}, 
done in collaboration with Silvia Mollerach and Esteban Roulet.

The IceCube collaboration first detected two PeV neutrino events~\cite{Aartsen:2013bka},
the highest energy neutrino events observed up till now.
After improving their sensitivity and extending their energy coverage down to
around 30~TeV, 26 additional events were observed~\cite{Aartsen:2013jdh}. 
With one year more of data this increased to a total of 37
events~\cite{Aartsen:2014gkd}.
These observations reject a purely atmospheric origin for all events at 
the 5.7$\sigma$ level. The best-fit $E^{-2}$ astrophysical spectrum 
with a
per-flavor normalization $(1:1:1)$ to these events suggests a flux level
of $E_{\nu}^2 \mathrm{d}\Phi_{\nu}/\mathrm{d}E = 10^{-8}$
GeV~cm$^{-2}$~s$^{-1}$~sr$^{-1}$ in the 100~TeV - PeV range.

To simulate the propagation of UHECRs and their secondary neutrinos and 
gamma rays, we used the publicly available simulation package 
CRPropa 2.0~\cite{Kampert:2012fi}, as was done in 
Ref.~\cite{Roulet:2012rv}. CRPropa includes all relevant 
interactions as well as cosmological and source evolution and redshift
scaling of the background light intensity in one dimensional (1D) 
simulations.

\section{Neutrino fluxes from UHECR protons}

For UHECR protons interacting with the CMB, the average neutrino energy from neutron 
decay for a typical neutrino production redshift of $z=1.2$ is $E_{\nu} \approx 6 \times 10^{15}$~eV,
while the average neutrino energy from the pion decay chain is  
$E_{\nu} \approx 10^{18}$~eV~\cite{Roulet:2012rv}. Wide peaks around these energies are expected
due to a wide $\Delta$ resonance, a wide thermal spectrum of CMB photons as
well as due to contributions from a wide range of redshifts. 

For UHECR protons interacting with the IRB, the neutrino energy from neutron
decay is typically $E_{\nu} < 10^{14}$~eV,
while the average neutrino energy from the pion decay chain is  
$E_{\nu} \approx 8 \times 10^{15}/[(1+z)(E_{\gamma}/\mathrm{eV})]$~eV,
with $R_{\mathrm{max}}$ the energy of the background photon.
The later is expected to give the dominant contribution to the neutrino flux
in the PeV range~\cite{Roulet:2012rv}.

Here we show the resulting spectra by simulating the propagation of
UHECR protons from their sources to the observer with CRPropa 2.0. 
We use 1D simulations including pair production, pion production and
all relevant decay channels. 
Furthermore, for all simulations, cosmological and source evolution
as well as redshift scaling of the background light intensity are included.
The IRB considered, including its redshift evolution,
is the 'best-fit model' of Ref.~\cite{Kneiske:2003tx}.
For this case a pure proton spectrum with a spectral index of
$\alpha = 2.4$, a minimum energy of $2 \times 10^{16}$~eV
 and a sharp cutoff at $E_{\mathrm{max}} = 200$~EeV 
has been injected at the sources. A continuous source density
following a redshift evolution (for the density times CR emissivity)
according to the gamma-ray burst evolution has been adopted.
This source evolution corresponds to the SFR6 model derived in Ref.~\cite{Le:2006pt}
and is here referred to as GRB2.

\begin{figure*}
  \centering
  \subfigure[Proton sources, $\alpha=2.4$, $E_{\mathrm{max}} = 200$~EeV, GRB2]{
    \includegraphics[width=0.47\textwidth]{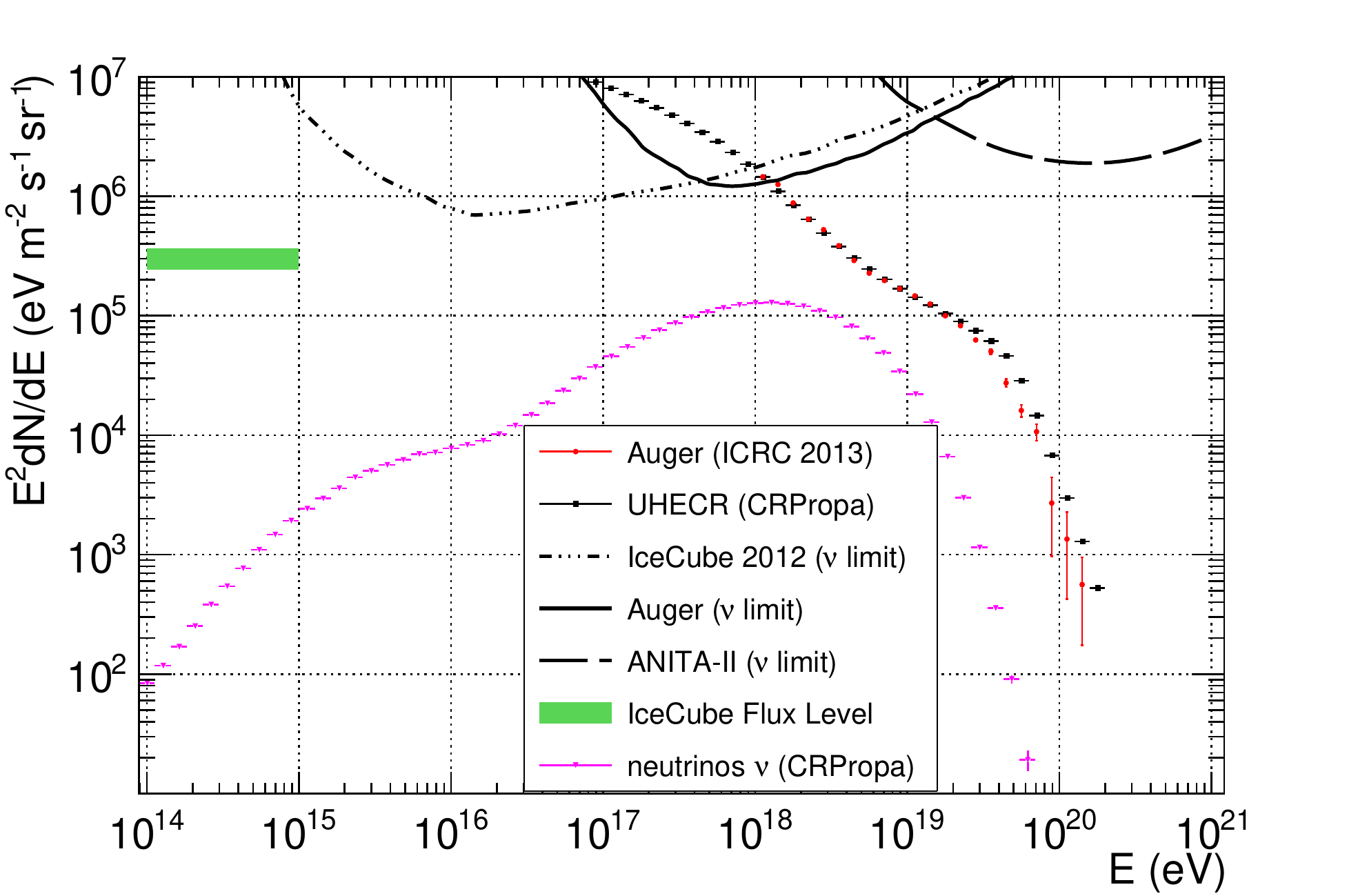}
    \label{fig:p_24_200_GRB2}}
  \subfigure[Including cascade photons]{
    \includegraphics[width=0.47\textwidth]{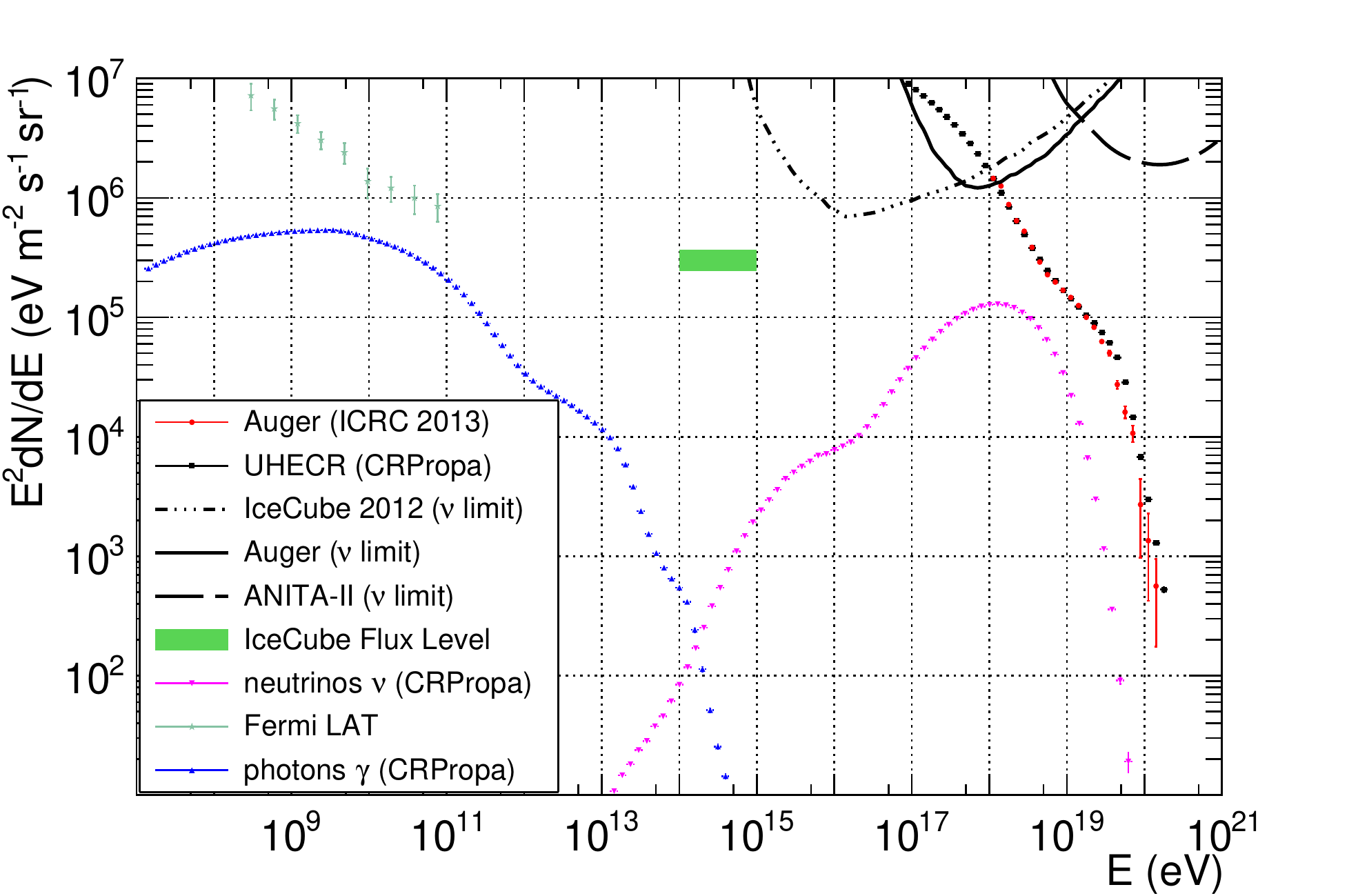}
    \label{fig:p_24_200_GRB2_combi}}
  \caption{Pure proton injection at the sources, with a spectral index at injection of 
	   $\alpha = 2.4$ and a maximum energy of $E_{\mathrm{max}} = 200$~EeV. The 
	    GRB2 source evolution model has been implemented. (a) In the left panel 
	    in red points the measured Pierre Auger UHECR spectrum is shown, while in black points
	    the simulated UHECR spectrum is given. The lines show the bounds on 
	    the all-flavor neutrino flux by IceCube (dashed dotted), Pierre Auger
	    (straight) and Anita (dashed). The green area indicates the flux level
	    of the IceCube events. The magenta points show the simulated neutrino flux. 
	    (b) The same spectra, bounds and flux level are given in
	    the right panel as well. Furthermore, the diffuse gamma-ray flux observed
	    by Fermi and the simulated gamma-ray flux from UHECR interactions are shown.
	    \label{fig:p_24_200_GRB2_total}}
\end{figure*}

The results are shown in fig.~\ref{fig:p_24_200_GRB2_total}.
The simulated CR spectrum has been normalized at 10 EeV to the spectrum
measured by the Pierre Auger Collaboration~\cite{Abraham:2010mj}
 as presented during the ICRC 2013~\cite{ThePierreAuger:2013eja}.
This spectrum has a 14\% systematic uncertainty on the energy scale,
which is not shown in the figures. The overall shape of the simulated
spectrum is in reasonable agreement with the measured spectrum.
The all-flavor neutrino spectrum has been normalized accordingly.
The bounds on the all-flavor neutrino flux obtained by IceCube~\cite{Aartsen:2013dsm},
Pierre Auger~\cite{ThePierreAuger:2013eja} and Anita~\cite{Gorham:2010kv} are displayed as well. 
It is clear that the simulated neutrino flux remains far removed from the neutrino bounds
as well as from the IceCube flux level.

Additionally, in fig.~\ref{fig:p_24_200_GRB2_combi}, the cascade photons originating from 
UHECR interactions with the CMB and IRB are shown. They nearly reach the diffuse 
gamma-ray flux level observed by Fermi~\cite{Abdo:2010nz}.

\begin{figure*} %[Proton sources, $\alpha=2.2$, $E_{\mathrm{max}} = 200$~EeV, FRII]
  \centering
  \includegraphics[width=0.94\textwidth]{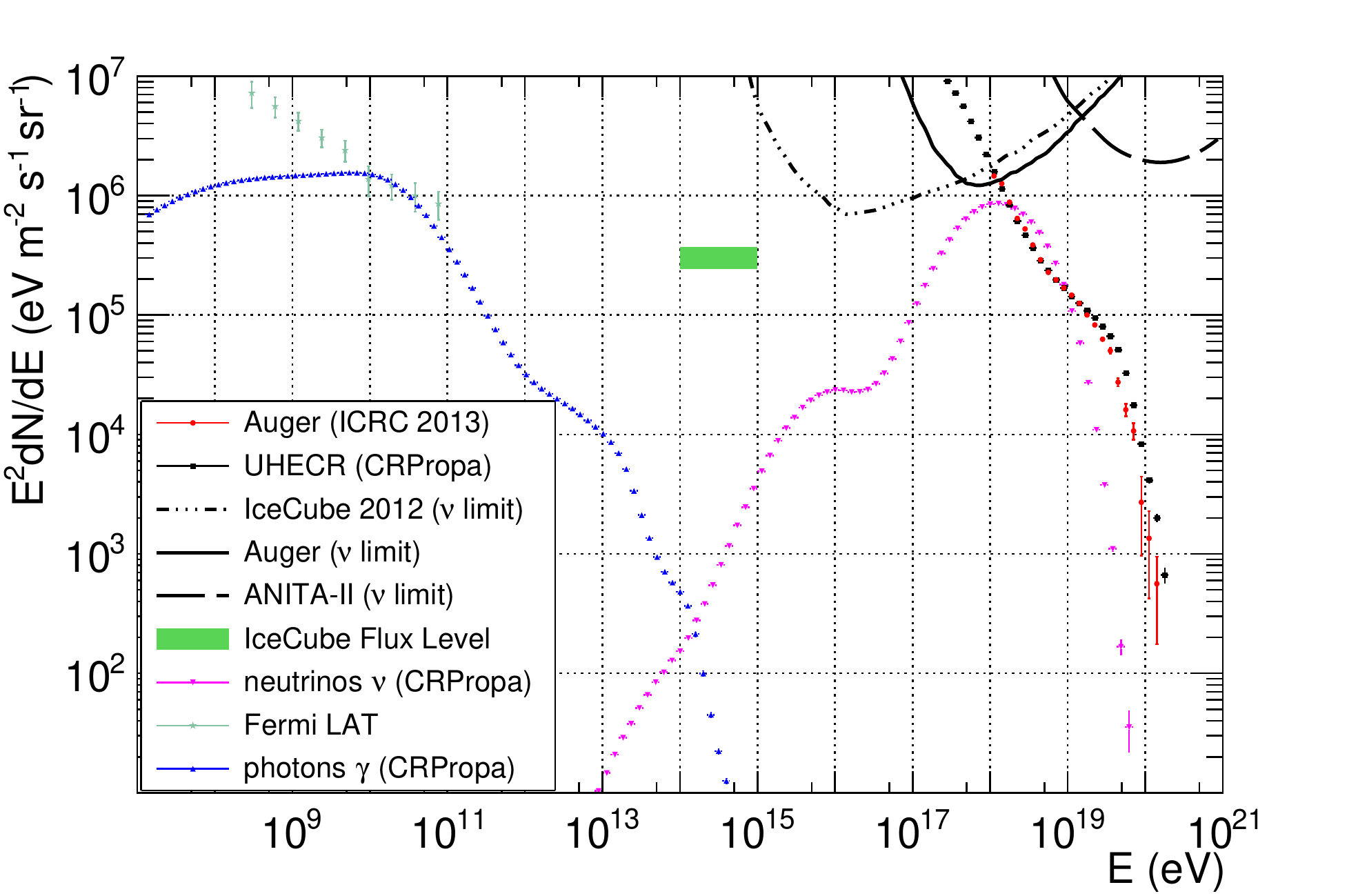}
  \caption{
  		Pure proton injection at the sources, with a spectral index at 
  		injection of $\alpha = 2.2$ and a maximum energy of 
  		$E_{\mathrm{max}} = 200$~EeV. The FRII source evolution
  		model has been implemented. 
  		The same simulated spectra, measurements, limits and flux level
  		are shown as in fig.~\ref{fig:p_24_200_GRB2_combi}.
	\label{fig:p_22_200_FRII}}
\end{figure*}

The neutrino flux is expected to increase when, 
instead of the GRB2 source evolution, a stronger source evolution as
for instance the AGN evolution model of Ref.~\cite{Wall:2004tg}
(here referred to as FRII) is implemented. This is confirmed by the simulation
results in fig.~\ref{fig:p_22_200_FRII}. In this case, instead of $\alpha = 2.4$, a
spectral index of $\alpha = 2.2$ at the sources was set, as this, for this scenario,
produces a closer resemblance to the measured UHECR spectrum. All other 
parameters have remained the same. 

However, as visible from fig.~\ref{fig:p_22_200_FRII}, not only the neutrino
flux but the photon flux increases with a stronger source evolution as well.
Whereas the simulated neutrino flux is still far removed from the IceCube flux level,
the gamma-ray flux is on the verge of conflicting with the diffuse gamma-ray
flux level observed by Fermi. 

Compared with Ref.~\cite{Roulet:2012rv} the Pierre Auger UHECR spectrum,
Pierre Auger neutrino limit, IceCube neutrino flux level and IceCube neutrino limit 
have been updated. Furthermore, for these simulations
an updated version of CRPropa 2.0 (CRPropa v2.0.4) was used, which includes
an improved energy interpolation for the pion production.

\section{Neutrino fluxes from iron nuclei}

When considering nuclei heavier than protons, photodisintegration can play an important role.
If a nuclei photodisintegrates completely, about half of the emitted nucleons will be
neutrons, which then decay to produce neutrinos. The energy of these neutrinos,
for interactions with the CMB background,
will typically be around $E_{\nu} \approx $~few $\times 10^{14}$~eV~\cite{Roulet:2012rv}.
Photopion production of nuclei on the CMB and IRB is possible as well, however
its threshold energy is a factor $A$ times higher than in the pure proton case, 
where $A$ is the mass number of 
the nucleus. This does give rise to the production of PeV neutrinos by photopion
production of IRB photons, but at a level which is expected to be lower than
that achievable in proton scenarios~\cite{Roulet:2012rv}. 

\begin{figure*} %[Iron sources, $\alpha=2.0$, $E_{\mathrm{max}} = 5200$~EeV, GRB2]
  \centering
  \includegraphics[width=0.94\textwidth]{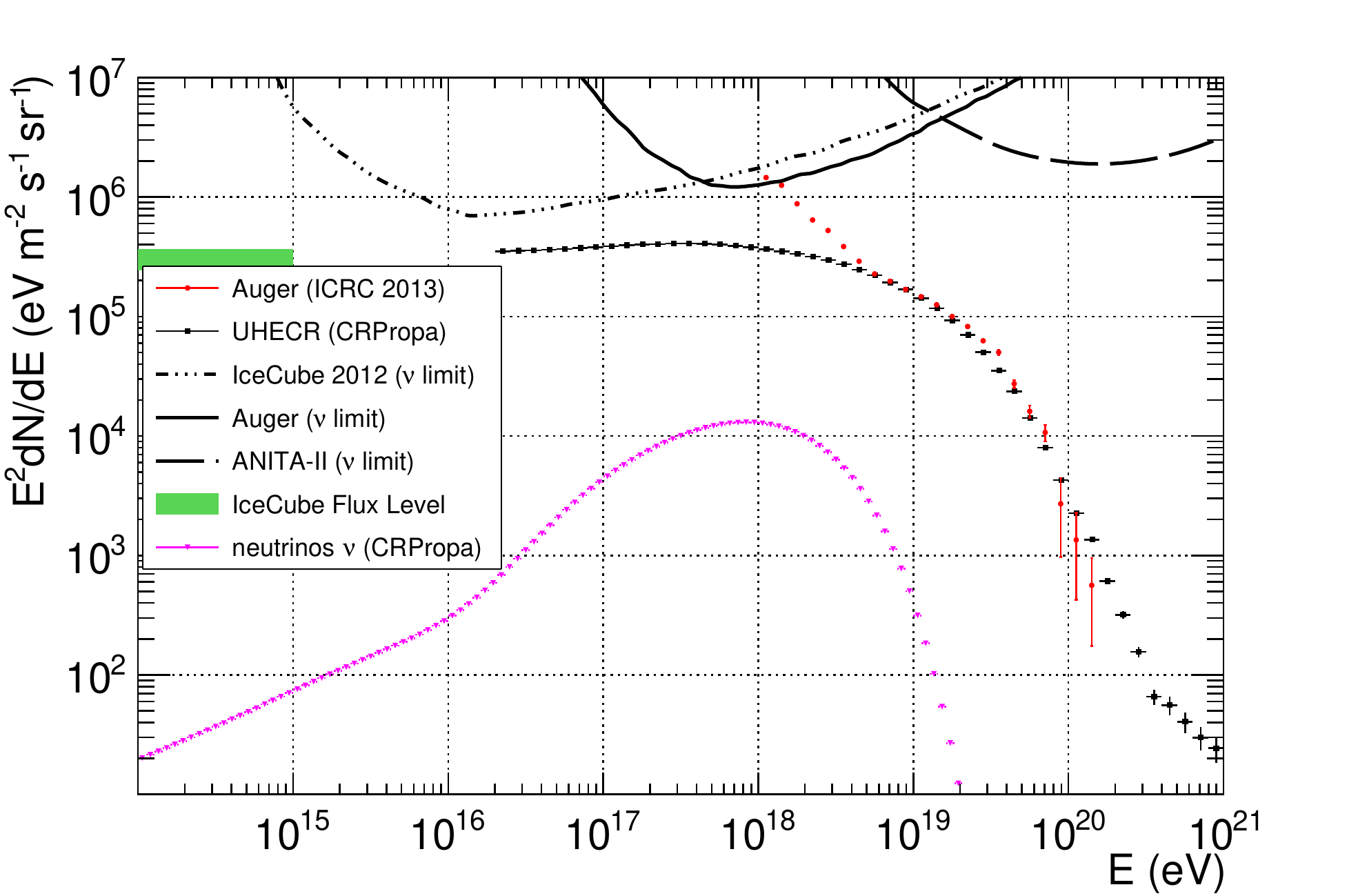}
  \caption{
		Pure iron injection at the sources, with a spectral index at 
		injection of $\alpha = 2.0$ and a maximum energy of iron of 
		$E_{\mathrm{max}} = 5200$~EeV. The GRB2 source evolution
		model has been implemented. 
		The same spectra, limits and flux level
		are shown as in fig.~\ref{fig:p_24_200_GRB2}. Compared with
		fig.~\ref{fig:p_24_200_GRB2} the neutrino flux has
		decreased due to the heavier composition.
	\label{fig:Fe_20_200_GRB2}}
\end{figure*}

This statement is confirmed by fig.~\ref{fig:Fe_20_200_GRB2}, which was obtained
by simulating pure iron injection at the sources, with a spectral index of $\alpha=2.0$
and a sharp cutoff at $E_{\mathrm{max}} = 5200$~EeV. In this case the GRB2 source 
evolution model was implemented. All other simulation parameters are the same as in the 
proton injection cases. The shape of the simulated spectrum is in reasonable agreement 
with the measured spectrum above the ankle.

\section{Neutrino flux for low $E_{\mathrm{max}}$ and mixed composition}

\begin{figure*} %[Proton and Iron sources, $\alpha=2.0$, $R_{\mathrm{max}} = 5$~EV, GRB2]
  \centering
  \includegraphics[width=0.94\textwidth]{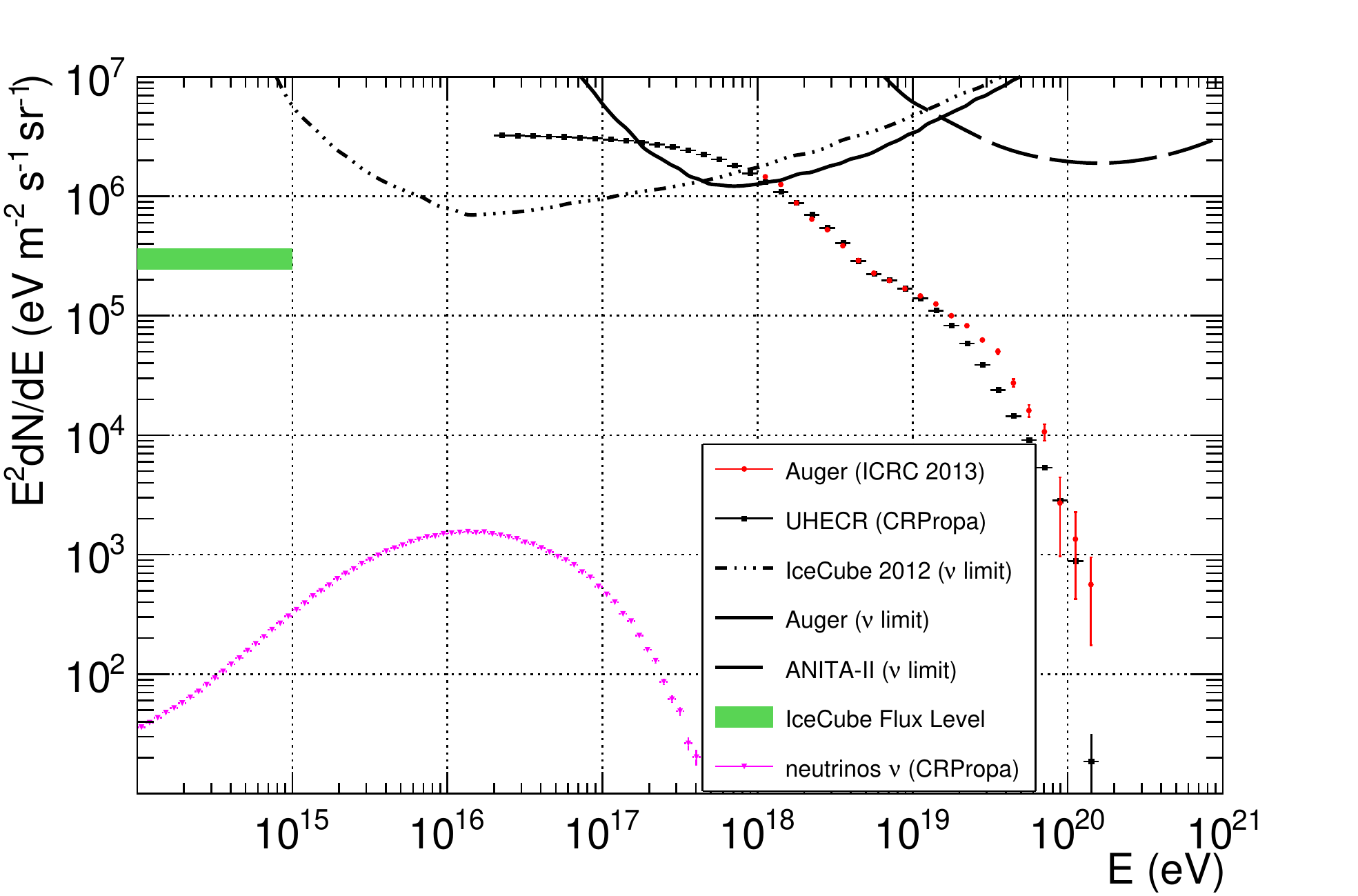}
  \caption{
	Proton and iron injection at the sources, with a ratio of 
	$n_p/n_{Fe}=250$ at a given $E/A$. The spectral index at injection
	is $\alpha = 2.0$ and the UHECRs are injected up to a maximum 
	rigidity of $R_{\mathrm{max}} = 5$~EV. The GRB2 source evolution
	model has been implemented. The same spectra, limits and flux level
	are shown as in the previous cases. The neutrino flux at EeV
	energies is reduced drastically due to the relatively low 
	$R_{\mathrm{max}}$, while at PeV energies it has
	increased compared to the pure iron case of fig.~\ref{fig:Fe_20_200_GRB2}
	due to the additional proton component.
	\label{fig:pFe_20_5_GRB2}}
\end{figure*}

Note that a lower maximum energy can drastically reduce
the neutrino peak at around $10^{18}$~eV, but is not expected
to significantly reduce the PeV neutrino flux for the pure iron
injection case. In fig.~\ref{fig:pFe_20_5_GRB2} a mixed-composition
 scenario is shown with proton and iron injected at the sources,
a spectral index at injection of $\alpha = 2.0$, a maximum rigidity of 
$R_{\mathrm{max}} = E_{\mathrm{max}}/Z = 5$~EV 
(with $Z$ the charge of the injected nucleus)
and the GRB2 source evolution model.
The ratio between injected proton and iron nuclei is $n_p/n_{Fe}=250$ 
at a given energy per nucleon $E/A$. It is clearly visible that the
EeV neutrino peak has been reduced drastically due to the low maximum
energy, while the neutrino flux at the PeV level has instead increased with respect to the
one in fig.~\ref{fig:Fe_20_200_GRB2} due to the additional proton primaries. 
In this case the shape of the simulated spectrum is in reasonable agreement 
with the full measured spectrum, while in the pure iron case it only
resembled the spectrum above the ankle.

\section{Conclusions}

For all the scenarios presented here the simulated neutrino flux 
remains at least one order of magnitude below the flux level indicated 
by the events measured by IceCube. When implementing stronger source
evolution models the expected neutrino flux can be enhanced. However,
when taking into account the secondary gamma-ray flux, it is clear that
the source evolution can not be enhanced too much in order not to 
exceed the diffuse gamma-ray flux observed by Fermi. Going to a heavier
composition than pure proton injection only decreases the neutrino
flux at the PeV level. So for all the scenarios considered here it is
difficult to interpret the IceCube events in terms of a cosmogenic 
neutrino flux, unless the IceCube events are a strong upward fluctuation
of the expected neutrino rates. 

\section*{Acknowledgments}

This work was supported by the Deutsche Forschungsgemeinschaft through
the collaborative research centre SFB 676, by BMBF under grants
05A11GU1 and 05A11PX1, and by the ``Helmholtz Alliance for
Astroparticle Physics (HAP)'' funded by the Initiative and Networking
Fund of the Helmholtz Association. GS acknowledges support from the 
State of Hamburg, through the Collaborative Research program 
``Connecting Particles with the Cosmos''.

\end{document}